\documentclass[11pt]{article}

\usepackage[utf8]{inputenc}
\usepackage{authblk}      
\usepackage[most]{tcolorbox}
\usepackage{float}
\usepackage{graphicx}
\usepackage{caption}
\usepackage{natbib}
\usepackage{url}
\usepackage{amsmath}
\usepackage{amssymb}

\title{Models of Ecological Fitting}

\author[1]{Rudy Arthur\thanks{Corresponding author: \texttt{R.Arthur@exeter.ac.uk}}}
\affil[1]{Department of Computer Science, University of Exeter, Exeter EX4 4PY, UK}
\author[2]{Arwen Nicholson}
\affil[2]{Department of Physics and Astronomy, University of Exeter, Exeter EX4 4QL, UK}
\author[3]{David M Wilkinson}
\affil[3]{Natural Sciences, University of Lincoln, UK and Classics and Archaeology, University of Nottingham, UK}

\date{\today}

\begin{document}

\maketitle

\begin{abstract}
 \noindent Ecological fitting describes a situation where species interact despite no shared evolutionary history. It is both a null model for adaptation and an important process in its own right, for example, the establishment of a non-native species is ecological fitting, and it is a major mechanism in the formation of Novel Ecosystems. In this paper we introduce and analyse a model of ecological fitting. We demonstrate both analytically and through simulation that the main factors determining invasion success are the invaded ecosystem's population and diversity, which set the size of the barrier for invaders. We also study the situation where many evolutionarily unrelated species simultaneously colonise an empty area. We show that a successful equilibrium is likely to be established, but the diversity tends to be very much less than the number of species in the starting pool and discuss the implications of this.
\end{abstract}

\section{Introduction}

The idea of ecological fitting was introduced by Janzen \citep{janzen1985ecological}. Inspired by observations in Santa Rosa National Park in Costa Rica, which, despite being mostly populated by species with very wide spatial distributions, still hosted myriad complex ecological interactions. As their wide spatial distribution meant these species and their interactions could not be evolutionarily adapted to each other and to that specific environment, he suggested that,
\begin{quote}
a major part of the earth's surface may be occupied largely by organisms that are rich in ecological interactions and have virtually no detailed evolutionary history with one another
\end{quote}\citep{janzen1985ecological}.

Ecological fitting built on Janzen's earlier observations \citep{janzen1980coevolution} pointing out that the mere occurrence of an interaction does not imply an evolutionary relationship. Thus, ecological fitting provides a null model for co-evolution \citep{ehrlich1964butterflies}. A co-evolutionary explanation of, say, a mutualism, would explain it as due to reciprocal selection. An ecological fitting explanation is that two independently evolved species whose ranges overlap and can happen to have mutually beneficial interactions.  While co-evolution certainly occurs, ecological fitting is a stronger baseline assumption than independence. Interactions are expected when species share a habitat, so co-evolution requires more evidence than just the observation of an interaction.

However, ecological fitting is more than a null model. As Agosta and collaborators have discussed \citep{agosta2006ecological, agosta2008ecological}, ecological fitting is important for understanding introduced species - including those that become `invasive'. As is well known, such species are successful and common \citep{van2018ecology,gioria2023invasive}, so ecological fitting is not a rare occurrence, nor are its effects small. An extreme example is provided by \citep{wilkinson2004parable}, who describes Ascension Island as ``a luxuriant tropical ecosystem constructed with almost no role for co-evolution.'' This system has become a classic example of what are now called Novel Ecosystems \citep{hobbs2013novel}.

A number of simulation models have been developed using ecological fitting to understand parasite host switching \citep{araujo2015understanding, feronato2022accidents} and ecological fitting is even operationalised to suggest effective strategies for species introduction, for example, \citep{le2017co} discuss the introduction of legume species into new areas, which must establish relationships with resident rhizobia. The success of such introductions can be enhanced by choosing species which are `promiscuous' i.e. interact with many other species, or by co-introduction of co-evolved rhizobia. Ecological fitting is also connected to other ideas in evolutionary biology, in particular as \cite{agosta2008ecological} argue, ecological fitting is an ecological example of exaptation \citep{gould1982exaptation}, where species interactions are not due to (co)evolution, but due to previously evolved traits of different species colliding in some new shared environment.

While there are mathematical models of ecological fitting, such as \citep{araujo2015understanding}, applied to specific cases, in this paper we introduce a more abstract framework. The aim is to study the idea of ecological fitting \emph{in general}. We are particularly interested in modelling the extreme example, inspired by  \cite{wilkinson2004parable}, where an entire ecosystem can be established by ecological fitting, which we will refer to as `ecological colonisation'. We believe this example highlights previously unrecognised but important ecological processes which may have consequences for how we understand evolution over geological time, as well as for nature restoration and recovery strategies.

In Section \ref{sec:description} we describe our modelling framework, which is built upon on standard ecological models. In Section \ref{sec:ecocolon} we describe how we use the model to study ecological colonisation and derive some general results for how the final diversity depends on the model parameters and the number of `seed' species. In Section \ref{sec:ecofit} we discuss invasion by ecological fitting, analysing the invasion barrier and how this depends on the population and diversity of the target ecosystem, as well as the properties of the invading species. In Section \ref{sec:discussion} we conclude and connect some of our modelling results to more practical issues in ecology.

\section{Model Description}\label{sec:description}

Our model will be based on the generalised Lotka-Volterra equation. Although there are a number of well known limitations of this well used model it ``remains irreplaceable in abstract theoretical  studies''\citep{rohr2025will}. The standard form of this equation is:
\begin{align}\label{eqn:lk}
    \frac{dN_i}{dt} = N_i\left( r_i   + \sum_{j} \alpha_{ij} N_j - \mu_i N_i \right)
\end{align}
giving the rate of change of the population $N_i$, of $M$ species labelled $i$, in terms of a growth rate, $r_i$, inverse carrying capacity, $\mu_i$, and interaction matrix, $\alpha_{ij}$. Solutions of the equation with random interaction matrices have been analysed extensively \citep{bunin2017ecological, servan2018coexistence, galla2018dynamically}. Generally these find that the equilibrium diversity $D$, i.e. the number of species with non-zero populations at late times, is an appreciable fraction of the species pool size, $M$ \citep{cornell2014species}.

We make two observations about Equation \ref{eqn:lk}. First, the carrying capacity term for species $i$, $-\mu_i N_i^2$, depends only on the population of $i$, not any other species. This implies that species do not impinge on each other. If each $i$ represents different tree species then in this model, the space, water, soil minerals etc. used by one species have no effect on the capacity of the environment to support other tree species. Second, while some species, like photosynthetic algae, can be said to have an intrinsic growth rate $r_i$, (at least under simplified laboratory conditions) many species, especially the macroscopic kind that typically interest ecologists, do not. Even plant `primary producers' rely on fungal associations and interactions with soil microbes and insects for key processes like nitrogen capture or pollination and are not viable without those species. Thus, allowing species to have an intrinsic growth rate may underestimate the importance of interspecies interactions.

We incorporate these observations into the Lotka-Volterra equation by setting $r_i=\sum_{j\neq i} J_{ij} \frac{N_j}{N}$ and $\alpha_{ij} = -\mu + \delta_{ij} \mu_i$ where $N$ is the total population $N = \sum_i N_i$. This gives
\begin{align}\label{eqn:lk2}
    \frac{dN_i}{dt} &=  N_i \left(\sum_{j \neq i} J_{ij} \frac{N_j}{N} - \mu N\right)
\end{align}
In this version, all species share a carrying capacity, so an increase in the population of one species gives less space for all other species. Not including self interaction, i.e. $J_{ii} = 0$, means no species is viable alone. Note the contribution of species $j$ to the growth of $i$ depends on the population fraction, $\frac{N_i}{N}$ and the total population is set by the value of $\mu$. 

This model represents the opposite extreme to the points made about the standard Lotka-Volterra equation. In reality some species, say trees, really do compete for space and light, so may be better represented by a shared carrying capacity. But a system of plants and birds should probably be modelled using separate carrying capacities, or, more accurately but also more difficult to model, the carrying capacity of an environment for birds depends on the number of trees. 

Equation \ref{eqn:lk2} is the model we study in this paper and represents the case we want to consider: species that must interact to survive while consuming a common resource. This means no species is viable alone, so inter-species interactions are crucial and we also have to consider the available carrying capacity, which can only support so many individuals. We choose $\mu = 0.01$, and set the entries of $J_{ij}$ as follows. With probability $\Theta$ set $J_{ij} = 0$, otherwise choose the values of $J_{ij}$ independently at random from a normal distribution with mean $0$ and variance $1$. $J_{ii} = 0$ always. 

We have previously investigated a version of this model with mutation through agent based simulations \citep{arthur2017entropic} where it is equivalent to the Tangled Nature Model \citep{christensen2002tangled}. Here we do not consider mutations and simply solve Equation \ref{eqn:lk2} numerically using scipy's \texttt{solve\_ivp} method \citep{2020SciPy-NMeth}. In the numerical solution we take the, ecologically justified, approach of setting to zero the population of any species that, at any point in the simulation, crosses below a threshold $\epsilon = 10^{-6}$. This accelerates convergence and represents the fact that extremely low populations are very vulnerable to extinction and their non-zero values are really artefacts caused by the finite running time of the simulations.

\section{Ecological Colonisation}\label{sec:ecocolon}

Our first model is based on a highly idealised version of the Green Mountain `experiment' described in \citep{wilkinson2004parable}. Briefly, Ascension is a relatively recently arisen volcanic island in the Atlantic. When first encountered it had few native species, for example it had around 25 native vascular plant species of which 10 were endemic. In the mid-19th century the British Admiralty began to rapidly introduce new plant species with the intention of producing a more benign environment for humans - so that today it is a `luxuriant tropical ecosystem' \citep{wilkinson2004parable} with at least 280 introduced vascular plant species. We will call the sudden and practically simultaneous arrival of multiple species in some area \textbf{ecological colonisation}.

\begin{figure}
    \centering
    \includegraphics[width=\linewidth]{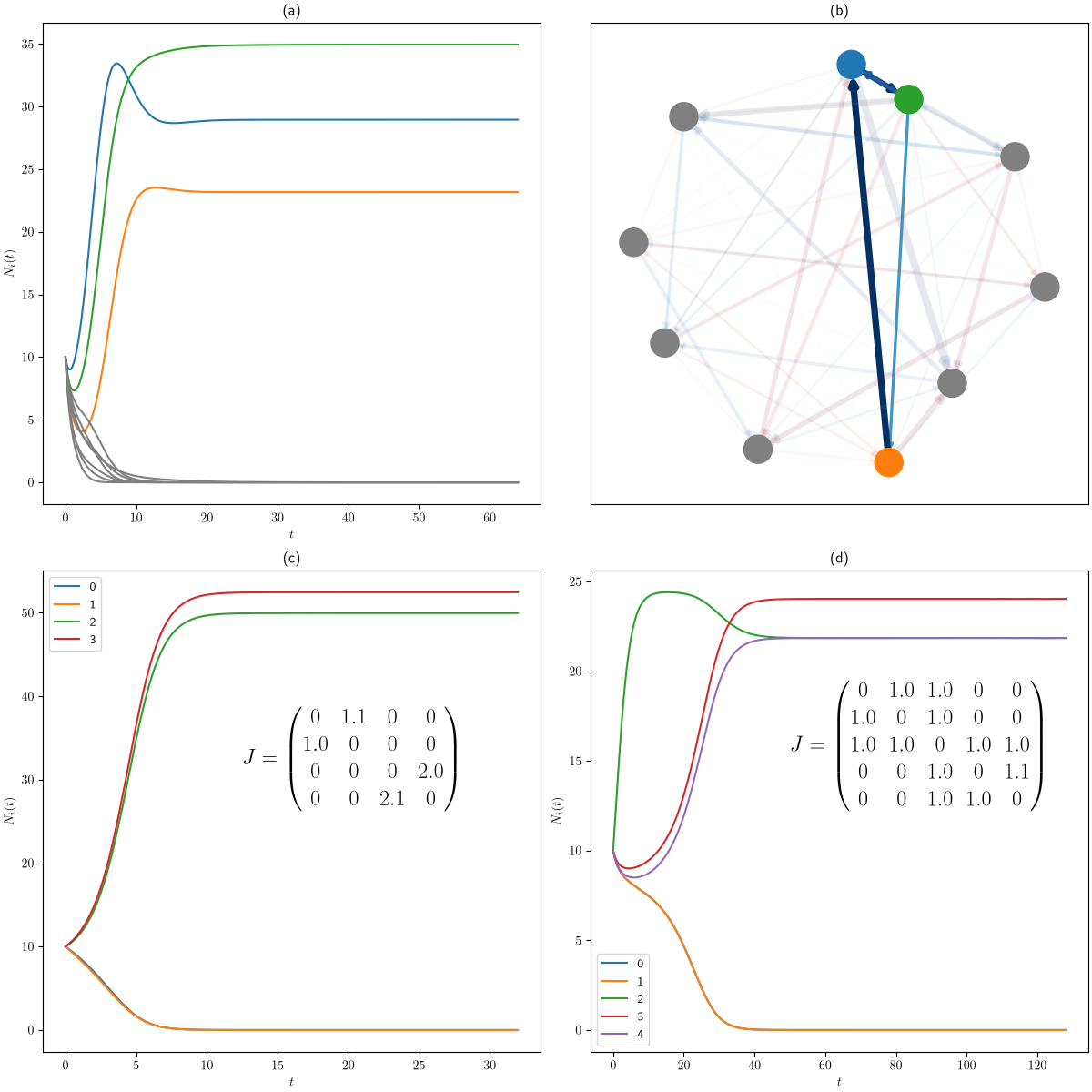}
    \caption{(a) Species populations $N_i$ starting with $M=10$ species and $\Theta = 0.25$. The grey species go extinct leaving $D=3$ at equilibrium. (b) The interaction matrix from (a) visualised as a network. The colours indicate the sign of the coupling, blue positive, red negative. The opacity indicates the size of $z_i J_{ij} z_j$ at the end of the simulation, the thickness indicates the magnitude of $J_{ij}$. (c) and (d) Show  populations simulated using the $J$ matrices shown starting with 10 individuals of each species.}
    \label{fig:1}
\end{figure}
For any choice of $M$, random values of $J_{ab}$ and some initial non-zero population of each species, in our case $N_i(0) = 10$, Equation \ref{eqn:lk2} rapidly converges to equilibrium. Some matrices $J$ do not have a non-zero equilibrium and yield final states with zero population, especially for small $M$. Otherwise, convergence is to a state where some subset of initial $M$ species have non-zero populations. An example is shown in Figure \ref{fig:1} (a) and (b). Stable oscillations are also possible, though we don't distinguish these from non-oscillating equilibria. Our question is, which species networks are present at equilibrium and why? 

To start answering this question, consider a system with 4 species present initially: $a$ and $b$ with $J_{ab} = J_{ba} > 0$ and $c$ and $d$ with $J_{cd} = J_{dc} > 0$ and all other interactions zero. Say $J_{ab} < J_{cd}$. Despite not being directly coupled $a,b$ and $c,d$ interact `indirectly' through competition for carrying capacity e.g. space or water. When all the populations are roughly equal at the start the fitness terms are
\begin{align*}
    f_a = f_b &\simeq \frac{ J_{ab} }{4} - \mu N\\
    f_c = f_d &\simeq \frac{ J_{cd} }{4} - \mu N\\
\end{align*}
so $f_a < f_c$ and the population of $a$ and $b$ will grow slower than $c$ and $d$ in some small time interval. This will change the ratios so $N_a, N_b < N_c, N_d$, and therefore $f_a,f_b$ will be even smaller than $f_c, f_d$ at the end of this small time increment. Ultimately, this leads to the extinction of $a$ and $b$. An example of such a trajectory is shown in Figure \ref{fig:1} (c). We set $J_{ab} \simeq J_{ba}$ and $J_{cd} \simeq J_{dc}$ so the different species can be distinguished in the plot. This shows that species can become extinct despite having no negative interactions with other extant species. The shared environment effectively introduces a negative interaction between every pair of species.

Generalising this example, if $J$ is block diagonal, summing Equation \ref{eqn:lk2} over a block $b$ gives
\begin{align*}
   \frac{d N_b}{dt} &= \sum_{i\in b} \sum_{j\in b} \frac{N_i J_{ij}N_j}{N} - \mu N N_b
\end{align*}
where $N_b = \sum_{i\in b}N_i$ is the block population. Define $n_i = \frac{N_i}{N_b}$ and $r_b = \sum_{i\in b} \sum_{j \in b} n_i J_{ij} n_j$ then
\begin{align*}
   \frac{d \log N_b}{dt} &=  \frac{N_b}{N}r_b - \mu N.
\end{align*}
Define $p_b = N_b/N$ as the ratio of the block population to the total. Comparing blocks $a$ and $b$ by subtracting their equations gives
\begin{align}\label{eqn:blockeqn}
  \frac{d}{dt} \log \frac{N_a}{N_b} =  p_a r_a - p_b r_b
\end{align}
implying that blocks with larger growth rate, $p_a r_a$, crowd out other blocks. Note this depends on the interaction sum, $r_a$, and the relative size of the block, $p_a$. Thus it is possible that different initial or intermediate population compositions could result in different final outcomes e.g. if the carrying capacity is `used up' by some block $a$ so $p_a \simeq 1$, it will be hard for some small group with $p_b \simeq 0$ to invade, even if $r_b$ is large.

In our experiments we will start with $N_i = 10$ as an initial condition. In this case $p_b = 1/|b|$ where $|b|$ is the number of species (diversity) of a block. Then
\begin{align}
  p_b r_b = \frac{|b|}{M} \frac{1}{|b|^2} \sum_{i\in b} \sum_{j \in b} J_{ij} 
\end{align}
So the block with the largest interaction sum realised by the fewest species will have the initial advantage.

In general, interaction matrices will not be block diagonal. Figure \ref{fig:1}(d) shows an example where species 2 is part of two mutually beneficial triples. The slight advantage of the lower one leads ultimately to the extinction of the first two species by crowding. Broadly speaking, species groups which have large mutual interactions are the ones which persist at equilibrium. 

As species are not viable alone, the lowest equilibrium diversity is $2$ and the highest is the starting diversity, $M$. There are $\binom{M}{2}$ possible species pairs, $\binom{M}{3}$ triples, $\binom{M}{4}$ quartets and in general $\binom{M}{D}$ possible subgroups with $D$ species. This is rapidly increasing up to $D = M/2$. Among the many potential groups, one will manage to win out. We call this process \textbf{virtual group selection}. Usually we consider group selection as acting between \emph{real} groups, perhaps different ant colonies or herds of zebras evading predators. In contrast virtual groups compete \textit{even while they overlap in members and are not spatially distinct.} 

\begin{figure}
    \centering
    \includegraphics[width=\linewidth]{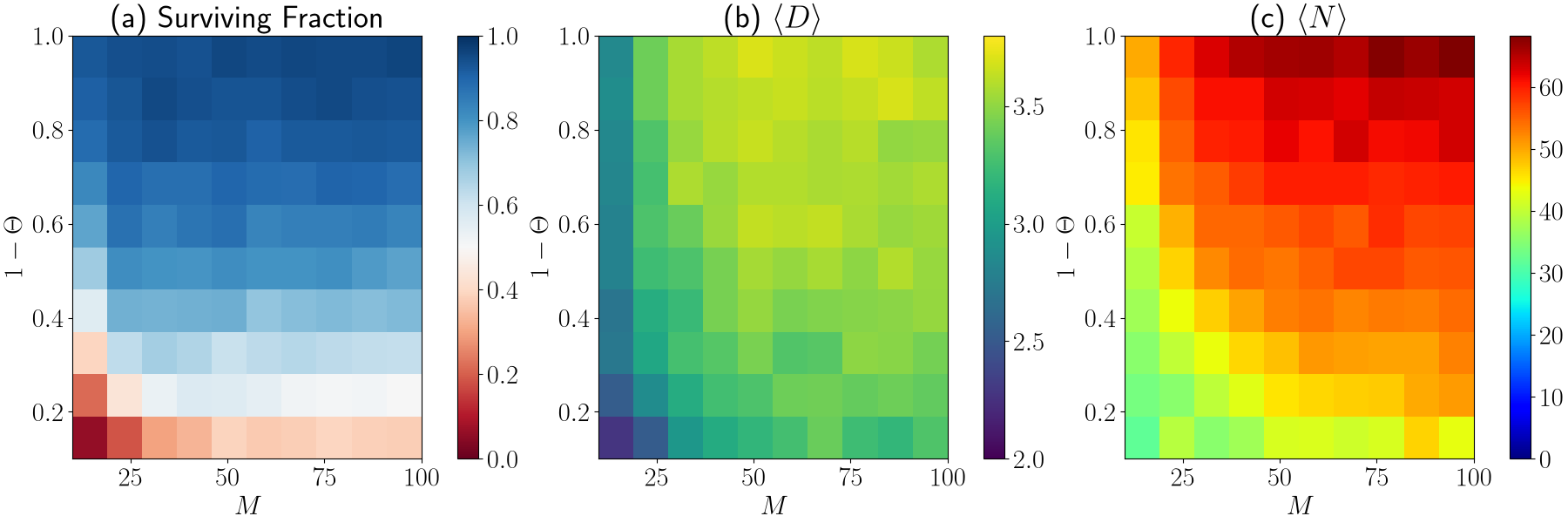}
    \caption{ Heatmap of average equilibrium state as a function of $1-\Theta$, connection probability and $M$, size of the species pool. Each square is the average over 1000 random $J$ matrices. (a) Shows the fraction of runs that end in total extinction. (b) Shows the average diversity in the non-extinct runs and (c) shows their average equilibrium population. }
    \label{fig:2}
\end{figure}
We give some heuristic arguments in Appendix \ref{sec:appA} for why we expect, despite the explosive growth in the number of possible groups of larger size, the winning group will tend to be be small, $D \ll M$ and the total population roughly $\frac{1}{\mu}\sqrt{(1-\Theta)(D-1)2\ln \frac{M}{D}}$. We test this by simulation in Figure \ref{fig:2} which shows ecological colonisation experiments for different values of $M$ and $\Theta$. Figure \ref{fig:2}(a) demonstrates that most of the runs find a populated stable state, where larger $M$ and more non-zero interactions makes this more likely. Figure \ref{fig:2}(b) shows the average final diversity $D$ is below 4, even for very large numbers of species in the starting pool. The maximum diversity (not shown) is never observed to exceed $10$, even for $M=100$ initial species. Figure \ref{fig:2}(c) shows that differences in the final populations are somewhat more, though still not very, sensitive to $M$ and $\Theta$.

In summary, analytical approximation and simulation of our model suggests that the final diversity observed in ecological colonisation experiments is much less than $M$, the number in the initial pool. We find that increasing $M$ and $(1-\Theta)$ is associated with slightly increased diversity and moderately more productive final states.

\section{Ecological Fitting}\label{sec:ecofit}
A less drastic and more common occurrence is the arrival of a non-native species in an established ecosystem. We refer to ecological fitting as the exaptation of that species to fit in its new habitat. Assume the invading species $i$ arrives with a small population so $N + N_i \simeq N$. From Equation \ref{eqn:lk2}, for the invader species to have positive growth rate requires
\begin{align}\label{eqn:barrier}
    \frac{dN_i}{dt} &> 0\\ \nonumber
    \implies \sum_{j \neq i} J_{ij} \frac{N_j}{N} &> \mu N
\end{align}
We refer to this as the \textbf{barrier equation}. Interactions of the new species with the pre-existing ones needs to be sufficiently positive to overcome the barrier set by $\mu N$ \citep{becker2014evolution}. 

Adding the new species has two main effects on the previous ecosystem. Any invader, if it grows significantly, increases $N$, reducing the available capacity, which can have a destabilising effect. New species may also have interactions with the previous ones, adding terms $J_{ai} \frac{N_i}{N}$. If $J_{ai} < 0$ the invader has an additional negative effect on $a$ other than crowding, further reducing the population of $a$ and possibly driving it to extinction. If $J_{ai} > 0$ then the invader benefits $a$, which could offset the increased crowding. 

\begin{figure}
    \centering
    \includegraphics[height=0.8\textheight]{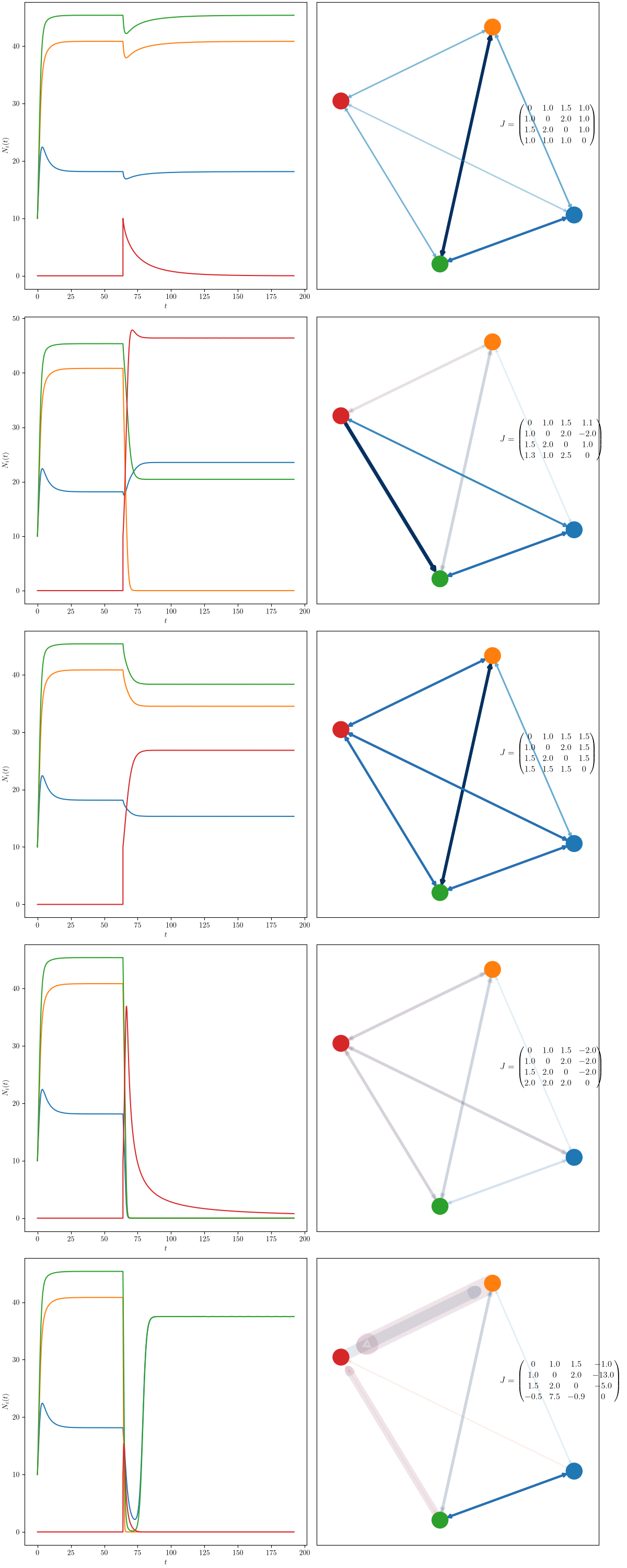}
    \caption{ Examples of invasions of the same system by different species showing 5 different outcomes. Top to bottom these are: invasion failure, displacement of species, addition, total extinction and disruption without establishment. Left shows the species' populations, right shows the interaction networks. }
    \label{fig:3}
\end{figure}
The outcome of an invasion depends on the balance of population ratios and interactions of the established ecosystem and the invader. In our simulations we have identified five possible outcomes due to the introduction of an invading species $i$:
\begin{enumerate}
    \item Failure of $i$ to establish.
    \item Successful invasion by displacement of one or more species.
    \item Successful invasion without displacement.
    \item Total extinction of the system.        
    \item Disruption of the equilibrium but no establishment of the invading species.
\end{enumerate}
 Examples of all five are shown in Figure \ref{fig:3}. Note that cases 2 and 3 can involve significant changes in the equilibrium populations of the previous species. Case 5 is quite rare, the example in Figure \ref{fig:3} required searching through several thousand random matrices and is quite sensitive to slight changes in the values of $J_{ij}$.

The barrier equation shows that large populations will tend to resist invasions. The current species fill the available space, an invader therefore needs to have very strong positive interactions with the extant species $r_i = \sum_{j \neq i} J_{ij} \frac{N_j}{N}$. Assuming for simplicity the population fractions are roughly equal, $N_j/N=1/D$, this means $r_i \sim \frac{1}{D}\sum_{j \neq i} J_{ij}$. This is a sum of $(1-\Theta)D$ independent normal random variables with mean 0 and variance 1, so $r_i \sim N(0,\frac{(1-\Theta)(D-1)}{D^2})$. 

This means high diversity ecosystems tend to resist invasion for two reasons. First, as shown in Figure \ref{fig:2}, high diversity is associated with higher populations, so we expect more diverse ecosystems to resist invasion due to an increase of the barrier, i.e. the term on the right hand side of Equation \ref{eqn:barrier}, $\mu N$. Second, since the variance of $r_i$ is, for large $D$, approximately $(1-\Theta)/D$ this means the potential for larger values of $r_i$ as $D$ increases is less, reducing the left hand side of the barrier Equation \ref{eqn:barrier}.  We also note that species with many interactions, $\Theta \simeq 0$, have a larger expected value of $r_i$, agreeing with the intuition that more promiscuous species are more likely to be successful invaders.

Rewriting the barrier equation in the equal ratio case, a successful invasion requires
\begin{align*}
     \sum_{j \neq i} J_{ij} > \mu N D 
\end{align*}
The left hand side is distributed normally with mean zero and standard deviation $\sqrt{(1-\Theta)D}$. Thus for large $D$ invasion success is quite unlikely. For example for $\Theta=0, D=4$, there is approximately a 2.5\% chance for an invader to have the required $2\sigma$ positive deviation from the mean. In general, for a normal distribution
\begin{align}\label{eqn:invasion}
P(r_i > \mu N D) &= 1 - \Phi\left(\mu N\sqrt{ \frac{D}{1-\Theta}}\right) \\ \nonumber
&\simeq \frac{1}{\mu N} \sqrt\frac{1-\Theta}{{2 \pi D}} \exp\left(-\frac{\mu^2 N^2 D}{2 (1-\Theta)}\right)
\end{align}
where $\Phi$ is the normal CDF, see Appendix \ref{sec:appA}. Thus the probability of overcoming the barrier is exponentially decreasing with increasing $D$ and super exponentially decreasing with increasing $N$.

\begin{figure}
    \centering
    \includegraphics[ width=\textwidth]{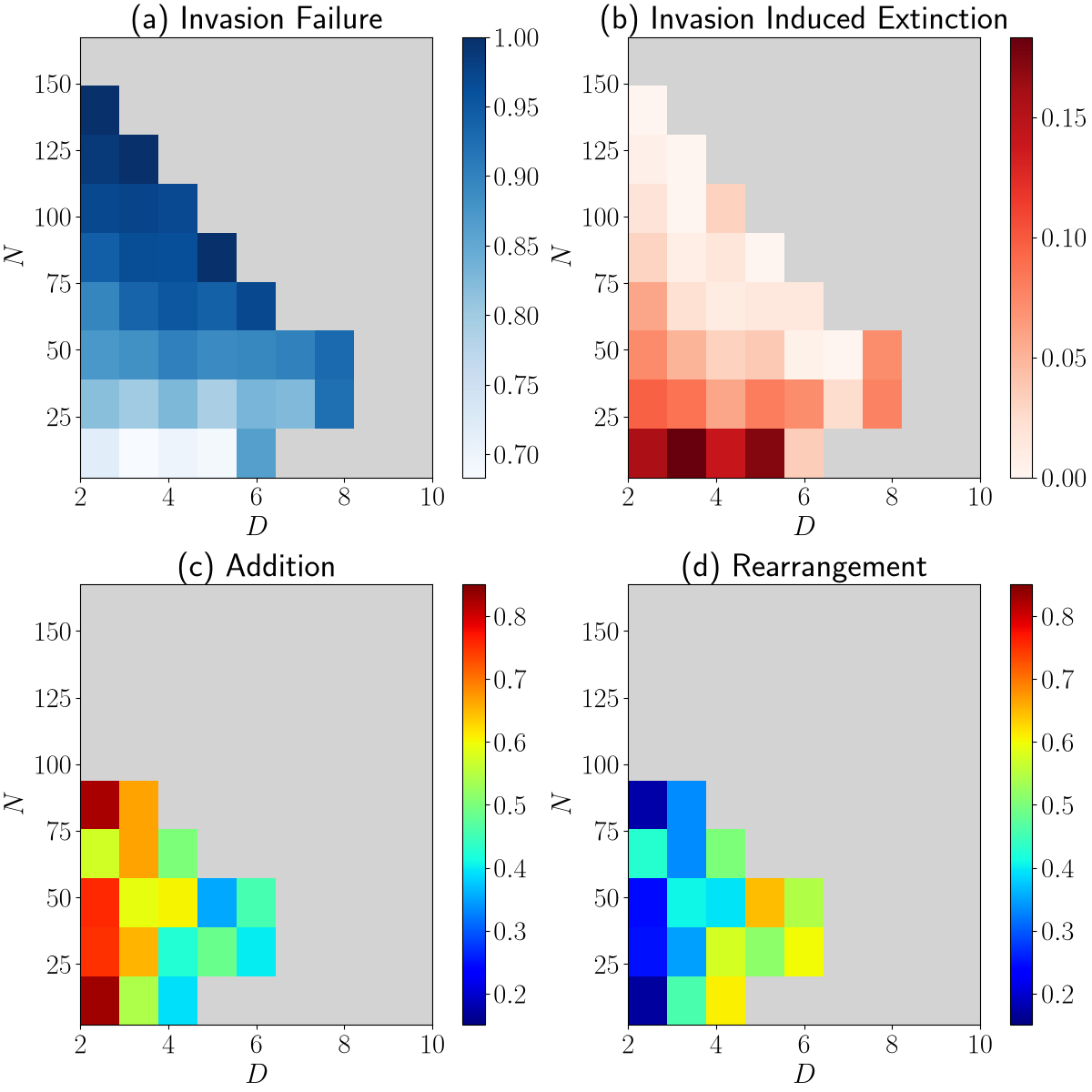}
    \caption{Binned fractions of invasion outcomes. Light grey indicates insufficient data. (a) Fraction of invasion failures (b) Fraction of invasion induced extinctions. Of the successful invasions, (c) is the fraction of additions and (d) the fraction of rearrangements. }
    \label{fig:4}
\end{figure}
We investigate this numerically in Figure \ref{fig:4}. We fix $\Theta=0.5$ and for each $M=10,20,\ldots,100$, perform 1000 simulations with different $J$ matrices to obtain equilibria with different values of $D$ and $N$. We then add an invading species $i$ with interactions chosen in the usual way with $\Theta=0.5$ and count the number of outcomes in each $N$ and $D$ bin. Only bins where at least 10 populated examples were found are shown.

Figure \ref{fig:4}(a) shows the fraction of failed invasions. This is quite high, approaching 1 for large $N$. The invasion probability is suppressed super-exponentially with $N$ ($e^{-N^2}$) so this is expected. Figure \ref{fig:4}(b) shows the fraction of invasions which result in total extinction. After invasion failure, this is the next most common outcome, especially for low diversity and low population states. 

Of course, ecologists mostly study ecosystems that are not extinct and study successful invasions rather than unsuccessful ones, an unavoidable sampling bias. We only consider successful invasions in the bottom row of Figure \ref{fig:4}. Figure \ref{fig:4}(c) shows, of the successful invasions, the fraction that resulted in adding the invader to the existing ecological network and Figure \ref{fig:4}(d) shows the fraction where the invader rearranges the ecological network, causing one or more species extinctions. We see that they are approximate mirrors of one another. If an invasion is successful, addition is favoured at low values of $D$ and rearrangement at high values. Case 5, invasion induced disruption without establishment of the invader, happens only once or twice across all experiments.

\section{Discussion}\label{sec:discussion}

Invasive species are well studied \citep{van2018ecology,gioria2023invasive}, and are a major practical problem for conservationists and field ecologists. If there is an empty niche a new species can of course fill it. Our model studies the more difficult case for the invader, when the ecosystem is full and resources more or less fully exploited. We ask the question if and when an invasive species can join or disrupt a stable state. Our model suggests, by theoretical consideration and numerical simulations, invasion is more likely to succeed at low population and diversity. If it does succeed, perhaps surprisingly, a common outcome is total extinction.

Invasion failure and total extinction are difficult to observe except in transition.  We note that many species know to have been introduced to Green Mountain on Ascension island did not successfully colonise \citep{duffey1964terrestrial}.   Restricting the discussion to cases where the invasion succeeds without causing total extinction, we find ecological fitting. When the invasive species has enough positive interactions with the extant species it can either join them or disrupt them. Probably the most interesting aspect of this invasion analysis is the notion of a barrier \citep{becker2014evolution}. From Equation \ref{eqn:invasion}, we see this barrier can be extremely difficult to overcome. Thus, for this model at least, where multiple species interact in a shared environment, the diversity-stability debate \citep{mccann2000diversity} is resolved. Higher diversity systems are more stable, in the sense that they better resist disruption by invasion.

The more dramatic example of ecological colonisation may seem initially to be of only limited interest; in unusual historical cases, like Green Mountain; for the construction of microcosms \citep{beyers2012ecological} or for sci-fi visions of terraforming \citep{nicholson2023there}. However we propose that something like ecological colonisation can actually be an important process when carried out repeatedly across geological time.

The mechanism of sequential selection \citep{lenton2018selection} proposes that periodic mass extinctions may explain long term planetary habitability. The version of \citep{arthur2022selection}, sequential selection with memory, is more explicit. A mass extinction causes not only a loss of species diversity, but also frees up a large amount of resources (carrying capacity). Mass extinctions, even ones as severe as the Permian-Triassic, are not total. The new world grows from the seeds of the old. Post extinction we have many species, present in low numbers, which compete to fill the now empty environmental niches. This is just like our ecological colonisation model and in particular, virtual group selection determines which species networks fill the now empty niches. As shown in Section \ref{sec:ecocolon} a larger species pool results in a more diverse and higher productivity ecosystem, though the effect is quite weak. This is in agreement with the findings in \citep{arthur2022selection} where it is argued in more detail that repeated global cycles of extinction and colonisation result in a \emph{slow} trend of increasing diversity and biomass.

A second, more practical, use of ecological colonisation on shorter and smaller scales, concerns recovery after disturbances, \emph{sensu} Grime \citep{grime2006plant}, such as wildfires, droughts and floods, all of which are becoming increasingly common as the climate changes. These catastrophes can cause substantial reductions in biomass, and potentially local mass extinctions. If the local extinction is complete enough, and occurs in a place where numerous species have overlapping ranges, virtual group selection may occur with the `winner' by no means guaranteed to have the same species members as the previous ecosystem. Especially vulnerable are species uniquely co-evolved to that particular habitat. Our models of ecological colonisation, as well as the practical example of Green Mountain \citep{wilkinson2004parable}, suggest we can still recover a productive ecosystem, but probably one lacking many of the unique species characteristic to particular locations.

Finally, ecological colonisation may have implications for nature restoration. Nature restoration involves returning species-poor agricultural land to something approaching a wilderness, often with minimal intervention \citep{carver2021guiding}, relying on dispersal and/or the expansion of small pockets of existing native species. This rapid expansion of a diverse species pool onto a `blank canvas' is quite similar to ecological colonisation. Since most nature restoration efforts use native species, and we do expect these to have co-evolved with species already present, the uncorrelated interaction matrix used here is overly pessimistic. 

However, in the modern world many species involved in less controlled experiments may not be native. For example, forests developing on abandoned agricultural land in Puerto Rico comprise a mix of native and many non-native tree species. \cite{lugo2004emerging} suggested that such new forests `will become increasingly prevalent in the biosphere in response to novel environmental conditions'.  Such novel systems are important as the are increasingly involved in providing ecosystem services, as more `natural' systems become less widespread. The Ascension Island Green Mountain system is an especially extreme case, but less extreme novel ecosystems are becoming increasingly common and important in applied ecology.

The results of Section \ref{sec:ecocolon} suggest that a high diversity species pool does not guarantee a high diversity outcome. Building on our model and exploring restoration through ecological colonisation models with more realistic interaction matrices will be interesting future work.

\section*{Acknowledgements}

We dedicate this paper to the memory of Paolo Sibani, who first introduced A.N. and R.A. to these kinds of models and whose kindness we will always remember.

\appendix

\section{Estimates of Equilibrium Diversity in the Colonisation Model}\label{sec:appA}

Given $M$ species, there are $\binom{M}{D}$ distinct groups of size $D$ which can be chosen. In a group of $D$ species with values $J_{ab}$ set as described in the text, there are, on average, $(1-\Theta) D(D-1)$ non-zero interactions. Analysing all possible groups analytically is tricky, because they are not independent. For example, at fixed $D$ the groups $\{1,2\}$ and $\{1,3\}$ are not independent, nor are different values of $D$ independent, e.g. groups $\{1,2\}$ and $\{1,2,3\}$. However we can make some simplifying assumptions and check against numerical simulation. Our goal is to get a sense of how many species to expect at equilibrium, given $M$ species in the initial pool.

A group $g$ of size $D$ will have a good chance of being selected if all of the internal interactions are positive. Although stable states with small negative couplings  can be found, negative values could cause one or more species to go extinct, reducing the group size. Therefore a reasonable, if simple, heuristic for likely values of $D$ is to estimate the chance of finding a set of $D$ mutually positive $J_{ab}$ values. As in the main text, all $J_{ab}$ are independent, taking the value 0 with probability $\Theta$, non-zero values are drawn independently from a standard normal and $J_{aa} = 0$ always. The chance that some value $J_{a\neq b} > 0$ is
\begin{align*}
    P(J_{a\neq b} > 0) &= (1-\Theta)\frac{1}{2} = q_+
\end{align*}
The probability of non-negative couplings is
\begin{align*}
    P(J_{a\neq b} \geq 0) &= \Theta + (1-\Theta)\frac{1}{2} =  (1+\Theta)\frac{1}{2} = q_0
\end{align*}
The chance that all non-diagonal values are positive is 
\begin{align*}
 q_+^{D(D-1)}
\end{align*}
with a similar formula for non-negative values. In general the chance that all values in a group are at least $c$ is
\begin{align}\label{eqn:qdecay}
 q_c^{D(D-1)}
\end{align}
where 
\begin{align}
    q_c = \Theta I(c \leq 0) + (1-\Theta)\Phi(c)
\end{align}
with $I(x)$ an indicator function, equal to 1 if its argument is true, otherwise equal to 0. $\Phi(c) = \frac{1}{2} \left( 1 + \text{erf}(\frac{c}{\sqrt{2}}) \right)$ is the CDF of the normal distribution. For any value of $q_c$ , Equation \ref{eqn:qdecay} is a rapidly decreasing function of $D$, so larger groups will most likely have some negative couplings, potentially destabilising them, resulting in lower realised values of $D$.

\begin{figure}
    \centering
    \includegraphics[width=\linewidth]{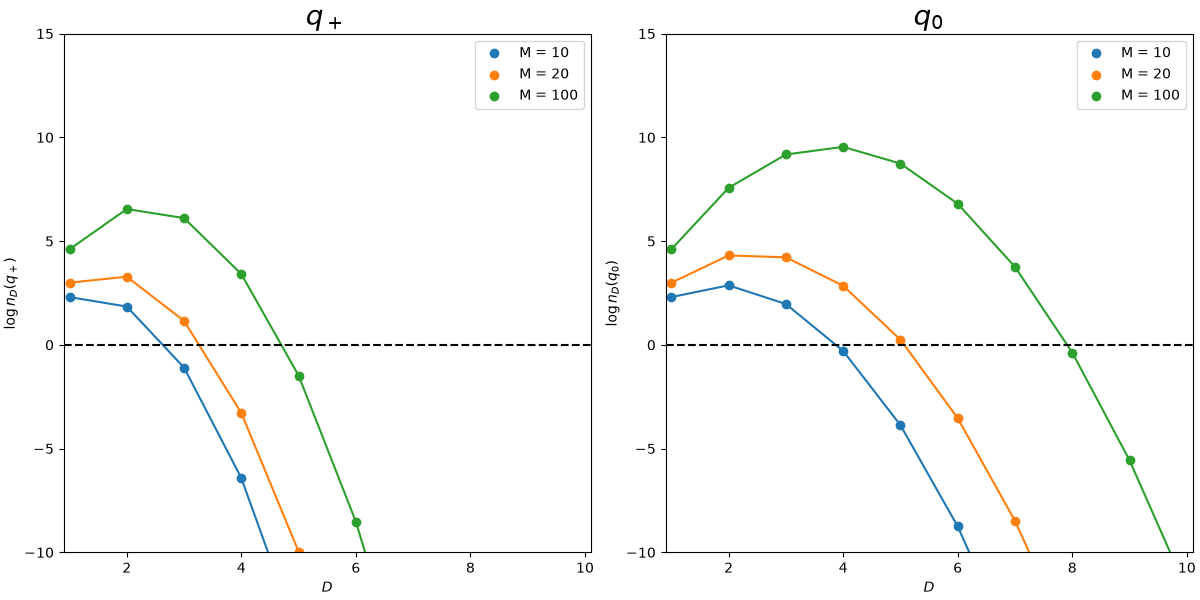}
    \caption{Log of the number of groups of of size $D$ with all positive couplings $n_D(q_c)$. Left uses $q_c = q_+$, all positive. Right uses $q_c = q_0$, non-negative. $\Theta = 0.25$.}
    \label{fig:app1}
\end{figure}
For some fixed size $D$ there are  $\binom{M}{D}$ possible groups and therefore roughly
$$
n_D(q_c) = \binom{M}{D} q_c^{D(D-1)}
$$
with all positive couplings. The power of $D^2$ eventually dominates, so $n_D \rightarrow 0$ at large $D$. The largest $D$ still giving at least 1 group is the solution to $n_D(q_c) = 1$. Using Stirling's approximation we can get an approximate analytical solution in terms of the Lambert W function, however Figure \ref{fig:app1} is more intuitive and shows that finding groups without any negative couplings is almost impossible at even moderate values of $D \ll M$.

Another way to analyse the groups is consider the sum of all the interactions in a group $g$ of size $D$. As in the main text, consider the simple case where the population ratios are equal,
\begin{align*} 
    \bar{J}_g = \frac{1}{D} \sum_{a \in g} \sum_{b\in g} J_{ab}
\end{align*}
This is a sum of, approximately,  $(1-\Theta) D(D-1)$ standard normal random variables. Since these are independent, $\bar{J}_g$ has distribution with mean $0$ and variance $\sigma_D^2 = (1-\Theta) (1-\frac{1}{D})$. Note this time we allow any number of negative couplings. At fixed $D$ the group that is likely to be selected is the group $g*$ where
\begin{align*}
   \bar{J}_{g*} = \max \left( \bar{J}_{g_1}, \bar{J}_{g_2}, \ldots, \bar{J}_{g_{K_D}} \right)
\end{align*}
$K_D = \binom{M}{D}$. We do not have independence of the $\bar{J}_g$s, however extreme value theory suggests if we did, a good approximation for the distribution of $\bar{J}_{g*}$ is a Gumbel distribution, which has the CDF
$$
G(x) = \exp \left( -\exp \left(-\frac{x - \alpha}{\beta} \right) \right)
$$
The location $\alpha$ and scale $\beta$ parameters can be estimated as \citep{gasull2015norming}
\begin{align*}
\alpha &= \sigma_D \left[ \Phi^{-1}\left(1 - \frac{1}{K_D}\right) \right]\\    
\beta &= \sigma_D \left[ \Phi^{-1}\left(1 - \frac{1}{eK_D}\right) - \alpha \right]
\end{align*}
where $\Phi^{-1}$ is the inverse of the normal cumulative distribution function, $\Phi^{-1}(x) = \sqrt{2} \text{erf}^{-1}(2x-1)$. This function is well-known from hypothesis testing e.g. $\Phi^{-1}(0.975) \simeq 1.96$ gives the familiar rule of thumb that a normal random variable has about a 5\% chance to differ from its mean by two standard deviations in either direction. The expected value of a Gumbel distribution is
$$
\mu_G = \alpha + \beta \gamma
$$
where $\gamma$ is the Euler-Mascheroni constant $\gamma = 0.57721 56649\ldots$. 

\begin{figure}
    \centering
    \includegraphics[width=\linewidth]{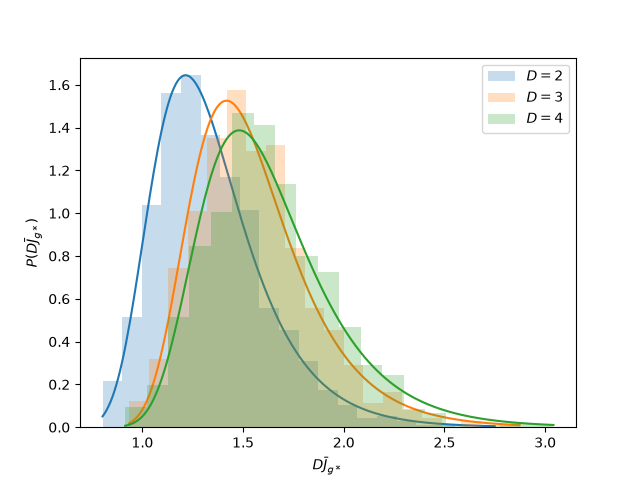}
    \caption{The histogram shows the values of $\bar{J}_{g*}$ obtained from 1000 random $J_{ab}$ matrices with $M=20, \Theta=0.25$. $\bar{J}_{g*}$ is found by computing $\bar{J}_{g}$ for \emph{all} groups of size $D=2, 3, 4$ for each generated matrix $J_{ab}$. The solid lines show the best fit Gumbel distribution. The expected values of $\bar{J}_{g*}$, to two significant figures are $1.21$, $1.42$ and $1.48$ for $D=2,3$ and $4$.  }
    \label{fig:app2}
\end{figure}
It is well known that the chance of large deviations from the mean is extremely low for a normal distribution, i.e. $\Phi^{-1}\left(1 - \frac{1}{K_D}\right)$ grows slowly as $D$ increases. Using the approximation \citep{gasull2015norming}
\begin{align*}
    \Phi^{-1}\left(1 - \frac{1}{K_D}\right) \simeq \sqrt{2 \ln K_D}
\end{align*}
we have
\begin{align*}
    \alpha &\simeq \sqrt{(1-\Theta) (1-\frac{1}{D})} \sqrt{2 \ln \binom{M}{D}} \\
    &\simeq \sqrt{(1-\Theta) (D-1) 2 \ln \frac{M}{D}}
\end{align*}
where the second step used Stirling's approximation for the binomial coefficient. Thus, despite the rapid growth of the number of groups with $M$, the maximum interaction sum grows much more slowly. 

The logarithmic growth of interaction strength (and thus total population) with the size of the starting pool $M$ agrees with  numerical results from the Tangled Nature Model \citep{arthur2022selection}. Figure \ref{fig:app2} shows some simulations, not relying on independence approximations, demonstrating the good fit of the Gumbel distribution and the slow increase in the expected value of $\bar{J}_{g*}$ with $D$.

In summary, finding groups without any negative interactions, which could destabilise the group and reduce $D$, is unlikely even for fairly small $D$. In addition, the analysis of the interaction sums shows even if we do find a large $D$ group, the payoff in terms of the potential growth rate, estimated by $\bar{J}_{g*}$ is not huge. Thus we don't expect to see groups with $D \simeq M/2$ selected, rather we expect only some fairly small fraction of the initial species pool to survive, $D \ll M$.

\bibliographystyle{plainnat}
\bibliography{references}

\end{document}